\documentclass[a4paper,10pt]{article}

\RequirePackage[frenchb,english]{babel}
\usepackage[T1]{fontenc}
\usepackage[latin1]{inputenc}
\usepackage{url}

\usepackage{graphicx}

\usepackage{latexsym}
\usepackage{epsfig}
\usepackage{amsmath}
\usepackage{times}
\usepackage{amssymb}
\usepackage{amsfonts}
\usepackage[mathscr]{eucal}

\newtheorem{theo}{Th\'eor\`eme}[section]
\newtheorem{prop}[theo]{Proposition}

\title{ \huge Approche poly\'edrale \\pour le probl\`eme du s\'eparateur \\ ~ \\ {\small M\'emoire du stage de Master Recherche \\ Juin 2006}}

\author{Marie-Jean MEURS}

\date{~}

\begin{document}

\maketitle

\begin{center}
\vspace{-1cm}
Sous la direction de Mohamed DIDI BIHA \\
\vspace{1cm}
Laboratoire Informatique d'Avignon\\
339, chemin des Meinajaries\\
Agroparc -- B.P. 1228\\
F-84911  Avignon Cedex 9\\
FRANCE\\
marie-jean.meurs(at)univ-avignon.fr\\
\end{center}

\vspace{1cm}



\abstract{
The vertex separator problem (VSP) in an undirected connected graph $G=(V,E)$ asks for a partition of $V$ into nonempty subsets $A$, $B$, $C$ such that $|C|$ is minimized subject to there is no edge between $A$ and $B$, and $max\{|A|,|B|\}\leq \beta (n) $ with $ \beta(n) $ integer such as $ 1 \leq \beta (n) \leq n$. We investigate the (VSP) in a polyhedral way, starting from the formulation given by E. Balas and C. De Souza \cite{Bal05a}. We introduce new efficient valid inequalities. Our algorithm is based on some of the inequalities discussed here. The power of these inequalities is illustrated by the computational tests, efficiently improving the Balas and De Souza's results \cite{Bal05b}.} \\

\begin{center}{ \textbf{R\'esum\'e}} 
\end{center}

Le probl\`eme du s\'eparateur (VSP) dans un graphe connexe non orient\'e $G=(V,E)$ est celui de la d\'etermination d'une partition de $V$ en trois classes non vides $A$, $B$, $C$ telle que $|C|$ soit minimum sous les contraintes qu'il n'existe aucune ar\^ete entre $A$ et $B$ et que $max\{|A|,|B|\} \leq \beta (n) $, o\`u $ \beta(n) $ est un entier tel que $ 1 \leq \beta (n) \leq n$. Le probl\`eme est \'etudi\'e d'un point de vue poly\'edrique \`a partir de la mod\'elisation de E. Balas et C. De Souza \cite{Bal05a} que nous avons enrichie. En particulier, nous introduisons de nouvelles in\'egalit\'es valides qui se r\'ev\`elent d'une grande efficacit\'e dans la r\'esolution de certaines instances difficiles du probl\`eme. Notre algorithme est bas\'e sur certaines des in\'egalit\'ees pr\'esent\'ees ici. Les r\'esultats obtenus am\'eliorent tr\`es sensiblement ceux de Balas et De Souza \cite{Bal05b}. \\

\noindent \textbf{Mots cl\'es :} VSP, graphe, s\'eparateur, polytope.\\



\section{Introduction}

Dans un graphe connexe non orient\'e, un s\'eparateur est un sous-ensemble de sommets ou d'ar\^etes dont la suppression d\'econnecte le graphe. La recherche d'un s\'eparateur de poids minimum \'equilibr\'e (dont la suppression isole deux sous-ensembles de sommets de tailles \'equivalentes, au plus $2n/3$) intervient dans de nombreux probl\`emes. 

Dans le domaine des r\'eseaux et t\'el\'ecommunications, un tel s\'eparateur est vu comme un goulot d'\'etranglement lorsque le r\'eseau est repr\'esent\'e par un graphe. Lors de l'\'evaluation de la capacit\'e du r\'eseau et de sa robustesse, il est utilis\'e pour la recherche de bornes inf\'erieures et l'identification des noeuds sensibles.

Dans le domaine des algorithmes op\'erant sur les graphes, la connaissance des s\'eparateurs de poids minimum \'equilibr\'es est fondamentale, en particulier pour ceux reposant sur le paradigme \og diviser et conqu\'erir \fg. 

Pour exemple, on peut citer les strat\'egies de r\'esolution des syst\`emes lin\'eaires creux de grande taille qui font appel \`a ce type d'algorithmes. La taille des probl\`emes impose une puissance de calcul croissante et le recours aux calculateurs parall\`eles est un passage oblig\'e. Ainsi, la factorisation des matrices creuses est particuli\`erement adapt\'ee \`a la programmation parall\`ele. Plus la matrice est creuse, plus l'on est susceptible d'avoir de parall\'elisme. Il faut conserver le plus possible l'aspect creux de la matrice durant la progression de la factorisation. Dans l'exemple de la factorisation de Cholesky, m\'ethode directe de r\'esolution des syst\`emes carr\'es, sym\'etriques d\'efinis positifs, la premi\`ere \'etape de la factorisation consiste en une renum\'erotation des lignes de la matrice initiale pour obtenir une matrice de Cholesky la plus creuse possible \cite{Erh93}. La recherche de la meilleure permutation est un probl\`eme complexe necessitant le recours \`a des heuristiques. Les algorithmes de renum\'erotation (degr\'e minimal, dissection emboit\'ee) reposent sur les principes de la th\'eorie des graphes. L'algorithme de dissection embo\^it\'ee repose sur le principe \og diviser et conqu\'erir\fg \ (\og\textit{divide and conquer}\fg). Il consid\`ere que la matrice initiale est irr\'eductible (i.e. le syst\`eme ne peut se d\'ecomposer en deux sous-syst\`emes ind\'ependants) et qu'elle est donc associ\'ee \`a un graphe d'intersection connexe \footnote{Le graphe d'intersection d'une matrice $A$, not\'e $G(A)$, a un sommet pour chaque colonne de $A$ et une ar\^ete entre une paire de sommets si le produit scalaire des colonnes correspondantes est non nul.}. Le but de la dissection embo\^it\'ee est de faire interagir le moins possible des parties relativement ind\'ependantes de la matrice. Pour cel\`a, en partant du graphe initial G, on cherche un s\'eparateur qui permettra la d\'ecomposition de G en deux sous-graphes qui pourront \^etre trait\'es en parall\`ele. La proc\'edure est appliqu\'ee r\'ecursivement aux sous-graphes. Le parall\'elisme mis en \'evidence est tributaire de la taille des s\'eparateurs : plus ils sont petits, meilleur est le parall\'elisme.
Cet algorithme se d\'ecline selon plusieurs variantes (par exemple, la dissection emboit\'ee spectrale (\textit{Spectral Nested Dissection}) (SND) \cite{Pot92}) mais repose toujours sur la recherche de \og bons \fg \ s\'eparateurs. 

Dans le domaine du traitement automatique de la langue \'ecrite, le concept de s\'eparateur est utilis\'e en classification par Berry \textit{et al} dans les grands graphes de termes creux \cite{Ber04}. Les sommets du graphe sont les termes \footnote{syntagmes nominaux de plusieurs mots susceptibles de d\'esigner, par leurs propri\'et\'es syntaxiques et grammaticales, un objet ou une notion du domaine.} du domaine consid\'er\'e. Les ar\^etes du graphe repr\'esentent les relations syntaxiques entre les termes. L'objectif est d'obtenir, \`a partir de corpus textuels, des classes susceptibles de repr\'esenter des th\'ematiques. On proc\`ede par classification hi\'erarchique \cite{San03} mais certaines classes restent trop grosses pour \^etre interpr\'etables par l'utilisateur. Elles sont alors d\'ecompos\'ees en utilisant des s\'eparateurs minimaux complets. La qualit\'e de la d\'ecomposition est li\'ee \`a la taille des s\'eparateurs et \`a l'\'equilibre entre la taille des sous-classes obtenues.

Dans le domaine de la bioinformatique, les s\'eparateurs sont recherch\'es dans les graphes de grille mod\'elisant la structure des prot\'eines. Fu \textit{et al} \cite{Fu05} explorent la r\'esolution du probl\`eme des configurations d'\'energie minimale qui doivent rassembler les sommets non cons\'ecutifs occupant des positions voisines. \\

Formellement, le probl\`eme du s\'eparateur (VSP) peut \^etre \'enonc\'e de la fa\c{c}on suivante :

Donn\'ees :
\begin{itemize}
	\item $G=(V,E)$ un graphe connexe non orient\'e avec $|V|= n$,
	\item $\beta (n) $ entier tel que $ 1 \leq \beta (n) \leq n$,
	\item $c_i$ co\^ut associ\'e \`a chaque sommet $ i \in V$.
\end{itemize}
\bigskip

Objectif :

Trouver une partition $ \{A,B,C \} $ de $V$ telle que : \\
\begin{align}
	& E \ \mathrm{ne \ contient \ aucune \ ar\hat{e}te \ (} i,j \mathrm{) \ avec} \ i \in A \ , j \in \ B, \medskip \label{condun}\\
	& \mathrm{max\{ |A|,|B| \} }\leq  \ \beta (n) \ , \medskip \label{condeux} \\
	& \displaystyle{\sum_{j \in C} c_j } \ \mathrm{est \ minimum} \medskip \label{condtrois} 
\end{align}

\begin{figure}[ht]
	\centerline{\includegraphics[width=12pc]{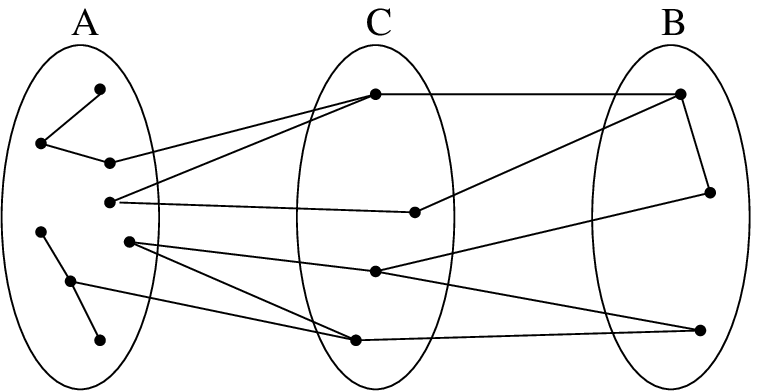}}
	\centerline{Exemple de s\'eparateur}
	\label{fig:abc}
\end{figure}
\bigskip

Le probl\`eme du s\'eparateur (VSP) est un probl\`eme NP-difficile \cite{Bui92}.

Le plus ancien r\'esultat concernant les s\'eparateurs est obtenu en 1869 par Jordan. Il d\'emontre alors que pour chaque arbre, un seul sommet le s\'epare en deux parties contenant chacune moins des deux tiers de ses sommets.
Lipton et Trajan \cite{Lip79} prouvent que tout graphe planaire poss\`ede un s\'eparateur de taille au plus $O( \sqrt n )$ qui le s\'epare en deux parties contenant chacune moins des deux tiers de ses sommets. Dans le cas o\`u $G$ est planaire et $\beta (n)=\frac{2n}{3}$, ils \'etablissent qu'un s\'eparateur de taille born\'ee par $ 2 \sqrt 2 \sqrt n $ peut \^etre trouv\'e en un temps $O(n)$.
Bui, Fukuyama et Jones \cite{Bui94} montrent que dans le cas des graphes planaires, la d\'etermination d'un s\'eparateur optimal est un probl\`eme NP-difficile.\\

En 2005, Egon Balas et Cid De Souza r\'ealisent la premi\`ere \'etude poly\'edrale du (VSP). Ils examinent le polytope des s\'eparateurs, enveloppe convexe des vecteurs d'incidence des partitions $\{A,B,C \} $ de $V$ v\'erifiant (\ref{condun}) et (\ref{condeux}). Leurs recherches sont centr\'ees sur la relation entre s\'eparateurs et dominants. (Un ensemble dominant S d'un graphe G est un sous-ensemble de sommets de G tel que chaque sommet de G est soit dans S, soit voisin d'un sommet de S). Ils donnent une formulation du probl\`eme \`a l'aide d'un  programme lin\'eaire mixte \cite{Bal05a} et d\'eveloppent parall\`element un algorithme de branch-and-cut bas\'e sur les in\'egalit\'es issues des relations entre s\'eparateurs et dominants \cite{Bal05b}. Les recherches de Balas et De Souza ont \'et\'e notre point de d\'epart.\\

Notre travail repose \'egalement sur une approche poly\'edrale du (VSP). La formulation de Balas et de Souza utilise des variables sur les sommets du graphe G. Nous avons adjoint plusieurs in\'egalit\'es valides \`a celles pr\'esent\'ees. D'une part, nous avons consid\'er\'e les cha\^ines entre chaque paire de sommets non ajacents du graphe G, ce qui nous a permis une premi\`ere minoration du cardinal de tout s\'eparateur associ\'e \`a G. D'autre part, nous avons reformul\'e le probl\`eme en rajoutant des variables sur les ar\^etes.\\

Dans la seconde partie de ce document, nous exposons l'approche poly\'edrale du probl\`eme. Nous \'etudions le cas particulier du VSP o\`u $a \in A$ et $b \in B$ sont fix\'es. La dimension du poly\`edre associ\'e est donn\'ee et des in\'egalit\'es valides sont introduites. La troisi\`eme partie d\'eveloppe une mod\'elisation du probl\`eme utilisant des variables sur les ar\^etes du graphe. Notre impl\'ementation et les r\'esultats que nous avons obtenus sont pr\'esent\'es dans la quatri\`eme partie. La derni\`ere partie expose nos conclusions et les perspectives de nos travaux.

\pagebreak

\section{Le poly\`edre des s\'eparateurs}

Pour un graphe $G=(V,E)$ donn\'e, on consid\`ere le cas particulier du VSP o\`u deux sommets $a$ et $b$ non adjacents sont donn\'es et l'on cherche une partition $ \{ A, B, C \}$ satisfaisant (\ref{condun}) et (\ref{condeux}) avec $a \in A$, $b \in B$ et telle que $|C|$ soit minimum. Ce cas particulier sera nomm\'e probl\`eme du \og ab-s\'eparateur \fg. Dans cette partie, on \'etudie le poly\`edre associ\'e au probl\`eme du ab-s\'eparateur. En effet, la r\'esolution du VSP se ram\`ene \`a la r\'esolution d'au plus $\frac{n(n-2)}{2}$ sous-probl\`emes de ce type. 

\noindent Les sommets $a$ et $b$ non adjacents \'etant fix\'es, le vecteur d'incidence d'une partition $ \{ A, B, C \}$ v\'erifiant (\ref{condun}) et (\ref{condeux}) avec $a \in A$ et $b \in B$ est donn\'e par $X \in \{0,1 \}^{2(n-2)}$ : 

\begin{small}
\begin{displaymath}
X = ( x_{1a} , ... , x_{(n-2)a} , x_{1b} , ... , x_{(n-2)b})
\end{displaymath}
o\`u 
\begin{displaymath}
x_{ia} = 1 \Leftrightarrow i \in A , \ x_{ib} = 1 \Leftrightarrow i \in B,  \ \forall i \in V \setminus \{a, b \} 
\end{displaymath}
\end{small}
	
\noindent On d\'esigne par $P_{ab}$ le poly\`edre associ\'e au probl\`eme du ab-s\'eparateur, d\'efini par : \\
$P_{ab}=Conv \{ X \in \mathbb{R}^{2(n-2)}, \exists$ un ab-s\'eparateur $\{A,B,C\}$ dont $X$ est un vecteur d'incidence $ \}$.

\noindent Si $X$ est un vecteur d'incidence d'un ab-s\'eparateur $\{A,B,C\}$, alors $X$ v\'erifie :\\
\begin{align}
x_{ia} + x_{jb} \leq 1 \qquad & \forall (i,j) \in E \label{arijInq} \\				
x_{ja} + x_{ib} \leq 1 \qquad & \forall (i,j) \in E \label{arjiInq} \\
x_{ia} + x_{ib} \leq 1 \qquad & \forall i \in V\setminus \{a,b \} \label{ainterbInq} \\
\displaystyle{\sum_{i=1}^{n-2} x_{ia}} \leq \beta(n) - 1 \qquad & \forall i \in V\setminus \{a,b \} \label{AborneInq} \\
\displaystyle{\sum_{i=1}^{n-2} x_{ib}} \leq \beta(n) - 1 \qquad & \forall i \in V\setminus \{a,b \} \label{BborneInq} \\
x_{ia} \in \{0,1\}\qquad & \forall i \in V\setminus \{a,b \} \label{Integeria} \\
x_{ib} \geq 0 \qquad & \forall i \in V\setminus \{a,b \}  \label{trivialeInq0}
\end{align}

Les contraintes (\ref{arijInq}) et (\ref{arjiInq}) sont valides du fait de l'absence d'ar\^etes entre les ensembles $A$ et $B$. La contrainte (\ref{ainterbInq}) est valide car un sommet ne peut appartenir \`a la fois \`a $A$ et \`a $B$. Les contraintes (\ref{AborneInq}) et (\ref{BborneInq}) sont valides car les ab-s\'eparateurs v\'erifient (\ref{condeux}). 

\noindent Balas et De Souza \cite{Bal05a} ont montr\'e que \\
$P_{ab}=Conv \left \{ X \in \mathbb{R}^{2(n-2)} \ , \ X \ v \acute{e} ri \!fie \ (\ref{arijInq}) \ \grave{a} \ (\ref{trivialeInq0}) \right \}$.

\subsection{Dimension du poly\`edre $P_{ab}$}

Un poly\`edre $P$ est de dimension $k$ si le nombre maximum de points affinement ind\'ependants de $P$ est $ k+1 $. Si $P \subseteq  \mathbb{R}^k $ alors $dim(P) \leq  k$.\\

\noindent Soient $V(a) \subset V$ et $V(b) \subset V$ les ensembles de sommets adjacents respectivement \`a $a$ et \`a $b$.
\begin{theo}{~}\label{th-dimp} \medskip \\
\noindent \mathversion{bold} $ dim (P_{ab}) = 2(n-2) - (|V(a)| + |V(b)|)$. \mathversion{normal}
\end{theo}

\noindent \underline{Preuve} :

\noindent $\star$ On obtient rapidement $ dim (P_{ab}) \leq  2(n-2) - (|V(a)| + |V(b)|)$. En effet, pour tout sommet $i$ de $V(a)$ (resp. de $V(b)$), on a $x_{ib} = 0, \ \forall X \in P_{ab}$ (resp. $x_{ia} = 0, \ \forall X \in P_{ab}$). La description minimale de $P_{ab}$ contient donc au moins $|V(a)|+|V(b)|$ \'egalit\'es, d'o\`u la majoration de $dim(P_{ab})$.

\noindent L'\'egalit\'e est obtenue par r\'ecurrence sur le nombre de sommets du graphe :

\noindent $\star$ Pour $n=3$, on a $P_{ab}= \{(0,0) \}$ d'o\`u  $dim(P_{ab})=0$. \\ Or \ $ 2(n-2) - (|V(a)| + |V(b)|) = 0$, l'\'egalit\'e est donc v\'erifi\'ee.

\noindent $\star$ Supposons l'\'egalit\'e v\'erifi\'ee pour tout graphe dont le nombre de sommets est inf\'erieur ou \'egal \`a $ n-1$ et soit $G=(V,E)$ tel que $|V|=n$, $ \{a,b \} \subset V$.

\noindent $\bullet$ Cas 1 : $\exists v_0 \in V \setminus \{a,b \}$ tel que $v_0 \notin V(a)\cup V(b)$. \\ On suppose sans perte de g\'en\'eralit\'e que $v_0$ porte le num\'ero $(n-2)$ dans $G$.\\
Consid\'erons $G'$, le sous-graphe de $G$ induit par $V' = V \setminus \{ v_0 \}$. Soit  $P'_{ab}$ le poly\`edre des ab-s\'eparateurs associ\'es au graphe $G'$.\\
D'apr\`es l'hypoth\`ese de r\'ecurrence, $ dim (P'_{ab}) = 2(n-3) - (|V'(a)| + |V'(b)|) = p$.\\ $P'_{ab}$ contient $p+1$ points affinement ind\'ependants or $O_{\mathbb{R}^{2(n-3)}} \in P'_{ab}$ d'o\`u $P'_{ab}$ contient $p$ points lin\'eairement ind\'ependants $X'_i=(x_{1a}^{'i},...,x_{(n-3)a}^{'i},x_{1b}^{'i},...,x_{(n-3)b}^{'i})$, $i \in \{ 1,...,p \}$.\\
Soient $X_1,..., X_p$ les $p$ points de $\mathbb{R}^{2(n-2)}$ d\'efinis par : \\ $X_i = (x_{1a}^{'i},...,x_{(n-3)a}^{'i},\underbrace{0}_{x_{(n-2) a}},x_{1b}^{'i},...,x_{(n-3)b}^{'i},\underbrace{0}_{x_{(n-2) b}}) \quad i \in \{ 1,...,p \}$.\\ Les $X_i$ sont lin\'eairement ind\'ependants par construction et appartiennent \`a $P_{ab}$.\\ Consid\'erons les partitions : \\ $\Pi_1 = \{ \{a,v_0 \}, \{b \}, V \setminus \{a,b,v_0 \}\}$ et $\Pi_2 = \{ \{a\}, \{b,v_0  \}, V \setminus \{a,b,v_0 \}\}$. \\ $\Pi_1$ et $\Pi_2$ v\'erifient (\ref{condun}) et (\ref{condeux}) car $v_0 \notin V(a) \cup V(b)$.
Soient $X_{p+1}$ et $X_{p+2}$ les vecteurs d'incidence associ\'es respectivement \`a $\Pi_1$ et $\Pi_2$. $X_{p+1}$ et $X_{p+2}$ appartiennent \`a $P_{ab}$. D'o\`u, $\{O_{\mathbb{R}^{2(n-2)}},X_1,...,X_{p+2} \}$ est une famille de $p+3$  points affinement ind\'ependants. Par cons\'equent, $dim (P_{ab}) \geq p+2$ et $p+2 = 2(n-2)-(|V'(a)| + |V'(b)|)= 2(n-2)-(|V(a)| + |V(b)|)$.La derni\`ere \'egalit\'e vient du fait que $V(a)=V'(a)$ et $V(b)=V'(b)$. On a donc $dim (P_{ab}) = 2(n-2) - (|V(a)| + |V(b)|)$.

\noindent $\bullet $ Cas 2 : $v_0 \in V(a)\cup V(b), \ \forall \ v_0 \in V \setminus \{a,b \}$.

$\circ $ cas 2.1 : $\exists v_0 \in V \setminus \{a,b \}$ tel que $v_0 \notin V(b)$.\\
De m\^eme que dans le cas 1, on peut construire $p$ points $X_i$ lin\'eairement ind\'ependants appartenant \`a $P_{ab}$, avc $p=2(n-3) - (|V'(A)|+|V'(b)|)$. Soit $X_{p+1}$ le vecteur d'incidence du ab-s\'eparateur $\{ \{a,v_0 \}, \{b \}, V \setminus \{a,b,v_0 \}\}$.
La famille $\{O_{\mathbb{R}^{2(n-2)}},X_1,...,X_{p+1} \}$ est une famille de $p+2$  points affinement ind\'ependants de $P_{ab}$. On a donc $dim(P_{ab}) \geq p+1$ et $p+1 = 2(n-3)-(|V'(a)| + |V'(b)|) +1 = 2(n-2) - (|V(a)| + |V(b)|)$ car $ |V(a)| = |V'(a)|+1$ et $|V(b)|=|V'(b)|$.\\ On a donc $dim (P_{ab}) = 2(n-2) - (|V(a)| + |V(b)|)$. \\
Par sym\'etrie, on obtient un r\'esultat identique si $v_0 \notin A$.

$\circ $ cas 2.2 : $V \setminus \{a,b \} = V(a) \cap V(b)$, i.e. tout point de $V \setminus \{a,b \}$ est adjacent \`a $a$ et \`a $b$. On a alors $dim (P_{ab}) = 0$ or $|V(a)| = n-2$ et $|V(b)| = n-2$ d'o\`u $2(n-2) - (|V(a)| + |V(b)|) = 0$ et $dim (P_{ab}) = 2(n-2) - (|V(a)| + |V(b)|)$.
$\square$
\subsection{In\'egalit\'es valides}
Dans cette partie, nous pr\'esentons tout d'abord de nouvelles in\'egalit\'es valides bas\'ees sur la structure des cha\^ines entre a et b. Nous nous int\'eressons ensuite aux in\'egalit\'es associ\'ees aux sous-graphes connexes de $G$.
\subsubsection{In\'egalit\'es de cha\^ines}
\begin{prop}{~}\label{abchInq}\\
Soit $\Gamma_{ab}$ une cha\^ine entre $a$ et $b$ et $I(\Gamma_{ab})$ l'ensemble des sommets internes \`a $\Gamma_{ab}$.\\ L'in\'egalit\'e  
$\displaystyle{\sum_{i \in I(\Gamma_{ab})}(x_{ia} + x_{ib})} \leq |I(\Gamma_{ab})| - 1$ est valide pour $P_{ab}$.
\end{prop}
	\centerline{\includegraphics[width=12pc]{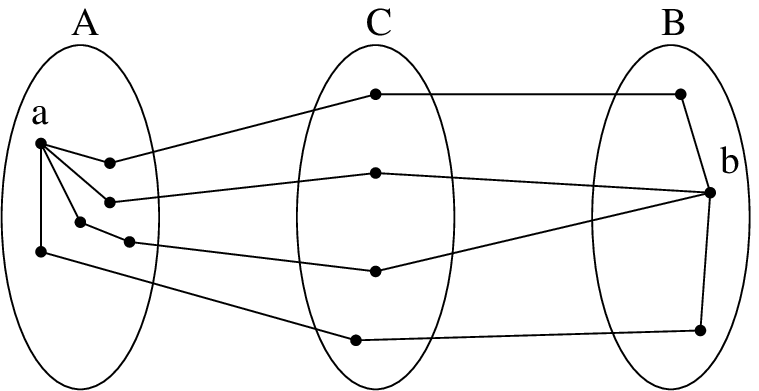}}
\noindent \underline{Preuve} :
La validit\'e de cette in\'egalit\'e vient du fait que toute cha\^ine entre $a$ et $b$ contient au moins un sommet appartenant \`a $C$ avec $\{A,B,C\}$ un ab-s\'eparateur \ tel que $a\in A$ et $b \in B$.


Pour toute paire $(i,j)$ de sommets non adjacents de $V$, soit $\alpha_{ij}$ le nombre maximum de cha\^ines sommets-disjoints entre $i$ et $j$. \footnote{Obtention de $\alpha_{ij}$: on construit un r\'eseau de transport, avec $i$ et $j$ pour entr\'ee et sortie, pour lequel un flot $\phi$ de valeur $\alpha_{ij}$ maximum entre $i$ et $j$ d\'etermine $\alpha_{ij}$ cha\^ines sommets-disjoints de $G=(V,E)$ entre $i$ et $j$. Pour cel\`a, on remplace dans $G$ chaque ar\^ete par deux arcs de directions oppos\'ees puis on d\'edouble tout sommet $x$ distinct e $i$ et $j$ par deux sommets $x'$ et $x''$ reli\'es par un arc $(x',x'')$ de capacit\'e 1. Tout arc qui allait vers $x$ est remplac\'e par un arc de capacit\'e infinie allant vers $x'$ et tout arc qui sortait de $x$ est remplac\'e par un arc de capacit\'e infinie sortant de $x''$. Le flot maximum entre $i$ et $j$ repr\'esente alors le nombre maximum de cha\^ines sommets-disjoints entre $i$ et $j$. \\
Pour chaque paire $(i,j)$ de sommets non adjacents, le flot maximum $\alpha_{ij}$ est obtenu \`a l'aide du code HI\_PR d'Andrew Goldberg disponible \`a l'adresse \url{http://www.avglab.com/andrew/soft.html}. Ce code est une version plus robuste du code PRF disponible \`a la m\^eme adresse. Il s'agit d'une impl\'ementation de la m\'ethode Push-Relabel pour les probl\`emes de flot maximum / coupe minimum (voir \cite{Gol97a} \cite{Gol97b}) qui est \`a notre connaissance la meilleure m\'ethode actuellement disponible pour r\'esoudre les probl\`emes de flot maximum.}\\
On d\'eduit imm\'ediatement de la proposition \ref{abchInq} que pour tout $\{A,B,C\}$ un ab-s\'eparateur \ tel que $a\in A$ et $b \in B$, le cardinal de $C$ est minor\'e par $\alpha_{ab}$.\\

%
%
%
%
%

Pour chaque paire de sommets $i$ et $j$ non adjacents de $V \setminus\{a,b\}$, les in\'egalit\'es suivantes sont valides pour $P_{ab}$ : \\
\begin{align}
\displaystyle{\sum_{k=1}^{n-2}(x_{ka}+x_{kb})} - \alpha_{ij}*(2-x_{ia}-x_{jb})\leq n - \alpha_{ij}  \label{ccij}\\
\displaystyle{\sum_{k=1}^{n-2}(x_{ka}+x_{kb})} - \alpha_{ij}*(2-x_{ja}-x_{ib})\leq n - \alpha_{ij}  \label{ccji}
\end{align}

\noindent Les in\'egalit\'es (\ref{ccij}) et (\ref{ccji}) traduisent le fait que si $\{A,B,C \}$ est un ab-s\'eparateur tel que l'un des sommets $i$ et $j$ est dans $A$ et l'autre dans $B$, alors $|C|$ est sup\'erieur \`a $\alpha_{ij}$.
%
\subsubsection{In\'egalit\'es associ\'ees aux sous-graphes connexes de $G=(V,E)$}
Soit $V'$ un sous-ensemble de $V$ tel que le graphe $G'$ induit par $V'$ est connexe et que $|V'| > \beta(n)$.\\
Soit $ \alpha_{0}^{V'}=Min\{\bar{\alpha}_{ij}, \ i \in V', \ j \in V', \ (i,j)\notin E \}$ o\`u $\bar{\alpha}_{ij}$ est le nombre maximum de cha\^ines sommets-disjoints entre $i$ et $j$ dans $G'$.

\begin{prop}{~}\label{connexInq}\\
L'in\'egalit\'e $\displaystyle{\sum_{i \in V'}(x_{ia} + x_{ib})} \leq |V'| - Min\{ \alpha_{0}^{V'} \ , \ |V'|-\beta(n) \}$ est valide pour $P_{ab}$.
\end{prop}

Cette in\'egalit\'e est int\'eressante dans le cas de graphes pour lesquels \\$\alpha_{min} = Min \{ \alpha_{ij}, \ i,j \in V ,\ (i,j) \notin E \}$ est petit mais dont certains sous-ensembles de sommets $V'$ induisent des sous-graphes connexes pour lesquels $ \alpha_{0}^{V'}$ est grand.

%
\section{Mod\`ele sur les ar\^etes}
\subsection{Formulation}
Le probl\`eme est formul\'e ici en utilisant des variables sur les ar\^etes du graphe $G =(V,E)$. On cherche donc un sous-ensemble d'ar\^etes $F \subseteq E$ tel que $F=F_1 \cup F_2$ avec $F_1 \cap F_2 = \emptyset$, pour lequel il existe une partition $\{A,B,C \}$ v\'erifiant (\ref{condun}) et (\ref{condeux}), $a \in A$, $b\in B$.\\$F$ est un sous-ensemble de $E$ ar\^etes-s\'eparateur. 

\noindent Le sous-graphe $(V_1 \cup V_2,F)$ induit par $F$, o\`u $V_1 = A\cup B$ et $V_2 = C$, est un graphe biparti. Les r\'esultats classiques pourront donc \^etre utilis\'es ici.

Le poly\`edre associ\'e au probl\`eme des sous-ensembles ar\^etes-s\'eparateurs est : \\
$Q_{ab} = Conv \{ \chi \in \mathbb{R}^{E}, \ \exists$ {\footnotesize un ar\^etes-s\'eparateur dont} $\chi$ {\footnotesize est le vecteur d'incidence} $\}$.

Pour toute cha\^ine $\Gamma$ , $\chi(\Gamma) = \displaystyle{\sum_{e \in \Gamma}} \chi(e)$. On note $\Gamma_{ab}$ une chaine entre a et b.

Si $\chi$ est vecteur d'incidence d'un ar\^etes-s\'eparateur F, alors $\chi$ v\'erifie les in\'egalit\'es suivantes :
	\begin{align}
	&\chi (\Gamma_{ab}) \geq 2 &\forall \ \Gamma_{ab} \label{2arInq}\\
	&\chi (\Gamma) - \chi (\Gamma_{ab}\setminus \Gamma) \leq |\Gamma| - 1  & \forall \ \Gamma_{ab} \quad avec \ \Gamma \subseteq  \Gamma_{ab}  \ et \ |\Gamma| \ est \ impair \label{impabInq}\\
	&\chi (\Gamma) - \chi (\Phi  \setminus \Gamma) \leq |\Gamma| - 1   &\forall \ \Phi  \ cycle \quad avec \ \Gamma \subseteq  \Phi  \ et \ |\Gamma| \ est \ impair \label{cyclimpInq}\\
	&\chi (e) \in \{0,1 \}  &\forall \ e \in E \label{arintegerInq}
	\end{align}
Les in\'egalit\'es (\ref{2arInq}) et (\ref{impabInq}) r\'esultent de l'appartenance de $a$ et $b$ \`a la m\^eme classe d'un graphe biparti : toutes les cha\^ines les reliant dans ce graphe sont paires et de longueur sup\'erieure ou \'egale \`a 2. L'in\'egalit\'e (\ref{cyclimpInq}) traduit l'absence de cycle impair dans un graphe biparti.
\begin{figure}[!h]
	\centerline{\includegraphics[width=12pc]{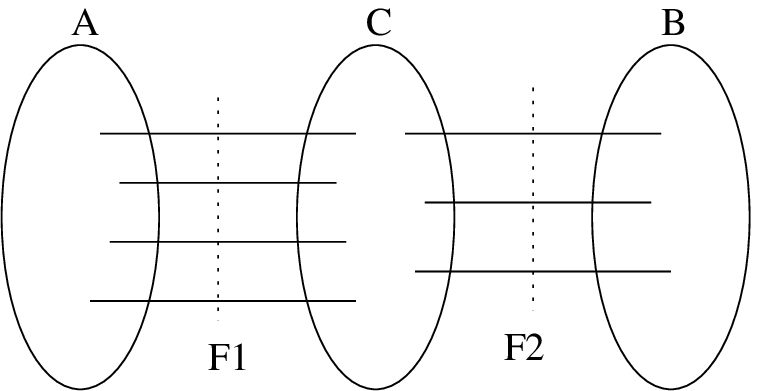}}
	\centerline{$F=F_1 \cup F_2$ un ar\^etes-s\'eparateur}
	\label{fig:aretes}
\end{figure}
\begin{prop}{~}\label{piab} \\
$Q_{ab}=Conv \{ \chi \in \mathbb{R}^{E} \ : \ \chi \ v \acute{e} ri \!fie \ (\ref{2arInq}) \ \grave{a} \ (\ref{arintegerInq}) \}$
\end{prop}

\noindent Soit $\bar{Q}_{ab}$ le poly\`edre d\'efini par : \\ $\bar{Q}_{ab}=Conv \{ \chi \ : \ \chi \ v \acute{e} ri \!fie \ (\ref{2arInq}) \ \grave{a} \  (\ref{cyclimpInq}) \ et \ 0 \leq \chi(e) \leq 1 , \ \forall e\in E\}$. \\ $\bar{Q}_{ab}$ est le poly\`edre issu de la relaxation du syst\`eme pr\'ec\'edent. \newpage On a imm\'ediatement $Q_{ab} \subseteq \bar{Q}_{ab}$ mais $Q_{ab} \not = \bar{Q}_{ab}$. En effet, l'exemple ci-dessous exhibe un point extr\^eme fractionnaire de $\bar{Q}_{ab}$ :\\ $\chi = (\frac{1}{2},1,1,\frac{1}{2},\frac{1}{2},1,1,\frac{1}{2},\frac{1}{2},1,1,\frac{1}{2})$.
\begin{figure}[!h]
	\centerline{\includegraphics[width=12pc]{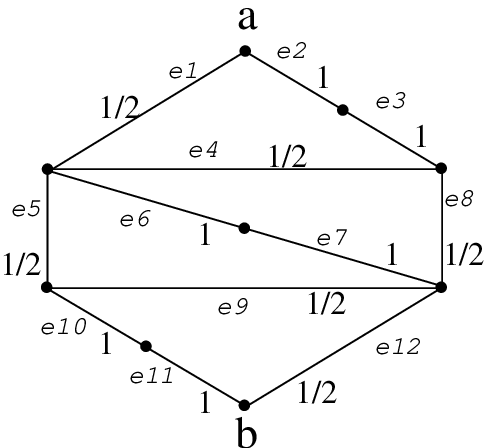}}
	\centerline{Point extr\^eme fractionnaire du syst\`eme relax\'e}
	\label{fig:relax}
\end{figure}

\noindent Ce point viole l'in\'egalit\'e \\ $ {\chi(e_1) + \chi(e_4) + \chi(e_8) + \chi(e_{12}) \geq \chi(e_2) + \chi(e_3) + \chi(e_6) + \chi(e_7) - 1}$ \\qui est valide pour $Q_{ab}$.

\section{Impl\'ementation et r\'esultats exp\'erimentaux}
\subsection{Mod\`ele}
\underline{Donn\'ees :}
\begin{itemize}  
	\item $G=(V,E)$ un graphe connexe non oriente 
	\item $\beta (n)$ entier tel que $ 1 \leq \beta (n) \leq n$, (ici, $\beta (n)=\frac{2n}{3}$ )
	\item $a$, $b$ affect\'es respectivement \`a $A$ et $B$ (ici $a$, $b$ sont des sommets fictifs)
	\item $ \alpha_{min} = \mathrm{Min} \{ \alpha_{ij}, \quad i \in V, \ j \in V, \ (i,j) \notin E  \} $
\end{itemize}
Sans perte de g\'en\'eralit\'e, on peut supposer que $|A| \leq |B|$. Comme on a \'egalement $|A|+|B| \leq n-\alpha_{min}$, il en r\'esulte $|A| \leq \lfloor \frac {n - \alpha_{min}}{2} \rfloor $, ce qui est traduit par (\ref{majA}) dans notre mod\'elisation.

\underline{Mod\'elisation :}\\
\begin{align}
\mathrm{Maximiser} \ \displaystyle{\sum_{i=1}^{n}(x_{ia} + x_{ib})} \notag \\
s.c. : \notag \\
x_{ia} \in \{ 0,1 \}  ,      &        \      & \forall i \in V  \label{xiaentier}\\  
x_{ia} + x_{ib} \geq 1  ,    &        \     & \forall i \in V \label{xia_xib} \\  
x_{ia} + x_{jb} \geq 1  ,    &        \      & \forall (i,j) \in E \label{xia_xjb} \\
x_{ja} + x_{ib} \geq 1  ,    &        \      & \forall (i,j) \in E \label{xja_xib}\\
\displaystyle{\sum_{i=1}^{n}(x_{ia} + x_{ib})}\leq n - \alpha_{min} \label{majC}\\	
1 \leq  \displaystyle{\sum_{i=1}^{n}x_{ia}} \leq  \lfloor \frac {n - \alpha_{min}}{2} \label{majA}\rfloor \\
1 \leq \displaystyle{\sum_{i=1}^{n}x_{ib}} \leq \beta (n) \ , \ 1 \leq \displaystyle{\sum_{i=1}^{n}x_{ia}} \leq \beta (n),\label{majBA}
\end{align}
\bigskip
\subsection{R\'esultats}
Nous avons travaill\'e exclusivement sur les instances \'etudi\'ees par Balas et De Souza pour disposer d'\'el\'ements de comparaison. Notre programme lin\'eaire mixte a \'et\'e r\'esolu en utilisant les APIs orient\'ees objet Concert Technology pour C++ du solveur Ilog Cplex 9.0. Tous les tests ont \'et\'e r\'ealis\'es sur un pc portable \'equip\'e d'un processeur Intel Pentium M 740 (fr\'equence 1,73 GHz) et de 1GB de RAM (DDR2). Balas et De Souza ont utilis\'e le solveur XPRESS dont les performances sont comparables \`a celles d'Ilog Cplex. En revanche, leurs r\'esultats sont obtenus sur un pc de bureau plus puissant que le n\^otre (Intel Pentium 4 \`a 2,5 GHz et 2GB de RAM).

Tous les graphes ont moins de 150 sommets except\'e \texttt{myciel7} qui en a 191. Pour chaque s\'erie d'instances, un tableau pr\'esent\'e en annexe rassemble les caract\'eristiques des graphes qui les composent. Ce tableau fournit le nombre de sommets de chaque graphe, le nombre d'ar\^etes, la densit\'e, la borne $\beta(n)$ associ\'ee et la valeur de $\frac{\alpha_{min}}{n}$ qui donne une premi\`ere indication sur l'efficacit\'e \'eventuelle de la minoration du cardinal de $C$ par $\alpha_{min}$.
  
\subsubsection{Instances DIMACS}
Les instances de cette cat\'egorie sont issues du challenge DIMACS sur la coloration de graphe et disponibles \`a l'adresse : \\
\url{http://mat.gsia.cmu.edu/COLOR/instances.html}\\
Nos r\'esultats sont pr\'esent\'es dans les tableaux qui suivent. La deuxi\`eme colonne indique le nombre de noeuds que nous avons obtenus, la troisi\`eme donnant la valeur correspondante pour Balas et De Souza. Les quatri\`eme et cinqui\`eme colonnes indiquent respectivement le temps n\'ecessaire \`a la r\'esolution de l'instance par notre algorithme et par celui de Balas et De Souza. L'ajout d'un ast\'erisque pr\'ecise que l'instance n'a pas \'et\'e r\'esolue \`a l'optimale par Balas et De Souza dans le temps imparti (1800 secondes). Balas et De Souza ont utilis\'e plusieurs m\'ethodes, nous avons syst\'ematiquement rapport\'e leur meilleur r\'esultat, le crit\`ere de choix \'etant le temps de calcul.

\begin{table}[!h]
\small
\centering
\caption{R\'esultats pour les instances DIMACS}
\label{resultdim}
\begin{tabular}{|l||r|r|r|r|} 
\hline
Instance     & \textit{nb. noeuds} & \textit{\scriptsize nb. noeuds B.S.} & \textit{temps(s)} & \textit{\scriptsize temps(s) B.S.}  \\ \hline \hline
david        &    113  &\scriptsize      52  &    0.984  &\scriptsize  0.19     \\\hline
DSJC125.1    & 132000  &\scriptsize  218098  &(1800)     &\scriptsize (1800) $\star$   \\\hline
DSJC125.1    & 523328  &\scriptsize  218098  & 9783.080  &\scriptsize (1800) $\star$   \\\hline
DSJC125.5    &    120  &\scriptsize   16768  &    8.610  &\scriptsize (1800)    \\\hline
DSJC125.9    &      0  &\scriptsize   34241  &    0.703  &\scriptsize   794.36  \\\hline
games120     &   8088  &\scriptsize   96920  &  121.047  &\scriptsize   429.02  \\\hline
miles500     &    145  &\scriptsize     318  &    7.156  &\scriptsize     2.11  \\\hline
miles750     &    938  &\scriptsize     369  &   62.797  &\scriptsize     9.83  \\\hline
miles1000    &    168  &\scriptsize      65  &   19.062  &\scriptsize    13.62  \\\hline
myciel3      &      0  &\scriptsize      19  &    0.000  &\scriptsize     0.00  \\\hline
myciel4      &     37  &\scriptsize      48  &    0.109  &\scriptsize     0.03  \\\hline
myciel5      &    290  &\scriptsize     169  &    0.828  &\scriptsize     0.28  \\\hline
myciel6      &    551  &\scriptsize     458  &   12.094  &\scriptsize     5.14  \\\hline
myciel7      &  14737  &\scriptsize    2441  &  880.875  &\scriptsize   160.59  \\\hline
queen6\_6    &      0  &\scriptsize      81  &    0.016  &\scriptsize     1.42  \\\hline
queen7\_7    &      0  &\scriptsize     263  &    0.063  &\scriptsize     7.78  \\\hline
queen8\_8    &      0  &\scriptsize    3533  &    0.031  &\scriptsize    42.44  \\\hline
queen8\_12   &   3175  &\scriptsize  162911  &   77.297  &\scriptsize (1800) $\star$   \\\hline
queen9\_9    &    959  &\scriptsize  291471  &   14.375  &\scriptsize  1067.5   \\\hline
queen10\_10  &   5412  &\scriptsize  126778  &  105.765  &\scriptsize (1800) $\star$   \\\hline
queen11\_11  &  13410  &\scriptsize   55220  &  456.781  &\scriptsize (1800) $\star$   \\\hline
queen12\_12  &  24635  &\scriptsize   25897  & 1245.080  &\scriptsize (1800) $\star$   \\\hline \hline
\textbf{moyenne}&\textbf{9751 }&\textbf{\scriptsize 49339}&\textbf{ 229.223}&\textbf{\scriptsize 634.97}\\\hline
\end{tabular}
\end{table}
En comparant nos r\'esultats \`a ceux de Balas et De Souza sur les instances DIMACS, nous obtenons :
\begin{itemize}
\item un nombre moyen de noeuds de 9751 contre 49339.
\item un temps moyen de r\'esolution de 229,223 secondes contre 634,97.
\item 20 solutions optimales (en moins de 30 minutes) contre 15.
\end{itemize}
On constate en particulier une tr\`es bonne efficacit\'e de notre m\'ethode sur les instances de types $n$-reines, pour lesquelles $\alpha_{min}$ est \'elev\'e, ainsi que sur \texttt{DSJC125.5} {\scriptsize ( $\alpha_{min}$=51)} et \texttt{DSJC125.9} {\scriptsize ($\alpha_{min}$=103)} r\'esolues \`a l'optimal respectivement en 8,61s {\scriptsize (120 noeuds)} et 0,703s {\scriptsize (0 noeuds)} alors que la premi\`ere n'est pas r\'esolue en moins de 30 minutes par Balas et De Souza et que la seconde est r\'esolue en 794,36s. Pour l'instance \texttt{DSJC125.1}, la meilleure valeur enti\`ere obtenue par Balas et De Souza en 30 minutes est 89 avec une borne sup\'erieure \`a 102,16. Nous obtenons dans le m\^eme temps une meilleure valeur enti\`ere de 91 avec une borne sup\'erieure \`a 101. Nous obtenons une solution optimale en 9783,080 secondes et la valeur de l'objectif est alors de 91, soit notre meilleure valeur enti\`ere apr\`es 30 minutes de calcul. ~\\~\\

\subsubsection{Instances MatrixMarket}
Cette cat\'egorie d'instances contient les graphes d'intersection de matrices de syst\`emes lin\'eaires de type $Ax=b$. Le graphe d'intersection d'une matrice $A$, not\'e $G(A)$, a un sommet pour chaque colonne de $A$ et une ar\^ete entre une paire de sommets si le produit scalaire des colonnes correspondantes est non nul. L'existence d'une ar\^ete $(i,j)$ dans $G(A)$ signifie donc qu'il existe une \'equation dans $Ax=b$ dans laquelle les deux variables $x_i$ et $x_j$ ont des coefficients non nuls. Ainsi, si le syst\`eme lin\'eaire est r\'esolu par une m\'ethode de type \og diviser pour r\'esoudre \fg \ (\og \textit{divide and conquer} \fg ), il pourra \^etre divis\'e en deux sous-syst\`emes plus petits r\'esolus s\'epar\'ement. La solution compl\`ete de ces sous-syst\`emes d\'epend des valeur des variables appartenant aux deux sous-syst\`emes. Le co\^ut algorithmique de la fusion des sous-syst\`emes n\'ecessaire \`a l'obtention de la solution du syst\`eme original augmente avec le nombre de variables communes aux sous-syst\`emes. L'efficacit\'e de l'algorithme impose aussi que la taille des sous-syst\`emes soit born\'ee par une fraction de la taille du syst\`eme initial. Le probl\`eme du choix de la meilleure fa\c{c}on de partitionner le syst\`eme lin\'eaire se ram\`ene donc \`a un VSP d\'efini sur $G(A)$. Les matrices utilis\'ees pour construire ces instances peuvent \^etre obtenues \`a l'adresse \url{http://math.nist.gov/MatrixMarket}. Elles correspondent \`a des probl\`emes pratiques issus de la physique, la m\'et\'eo, l'\'economie, etc. A partir de ces matrices, Balas et De Souza ont construit trois cat\'egories d'instances, MM-I, MM-II et MM-HD, selon les propri\'et\'es des matrices d'origine (voir \cite{Bal05b}). Elles peuvent \^etre obtenues \`a l'adresse : \\ \url{http://www.ic.unicamp.br/~cid/Problem-instances/VSP.html}

\newpage
Les graphes de la cat\'egorie MM-I ont au moins $20$ et au plus $100$ sommets. Leur densit\'e varie de $0,06$ \`a $0,67$ 
\begin{table}[!h]
\small
\centering
\caption{R\'esultats pour les instances MatrixMarket MM\_I}
\label{resultmm1}
\begin{tabular}{|l||r|r|r|r|} 
\hline
Instance     & \textit{nb. noeuds} & \textit{\scriptsize nb. noeuds B.S.} & \textit{temps(s)} & \textit{\scriptsize temps(s) B.S.} \\ \hline \hline
ash219     &  137 &\scriptsize    111 &   1.187 &\scriptsize    0.18  \\\hline
dwt72      &   26 &\scriptsize     59 &   0.500 &\scriptsize    0.07  \\\hline
can62      &   57 &\scriptsize    119 &   0.500 &\scriptsize    0.15  \\\hline
dwt66      &    8 &\scriptsize     31 &   0.312 &\scriptsize    0.06  \\\hline
bcspwr02   &   49 &\scriptsize     51 &   0.484 &\scriptsize    0.06  \\\hline
dwt\_\_59  &  137 &\scriptsize    201 &   0.969 &\scriptsize    0.30  \\\hline
bcspwr01   &   13 &\scriptsize     13 &   0.125 &\scriptsize    0.02  \\\hline
ash85      &  593 &\scriptsize    971 &   8.610 &\scriptsize    2.87  \\\hline
dwt87      &  109 &\scriptsize    241 &   3.031 &\scriptsize    0.90  \\\hline
impcol\_b  &  428 &\scriptsize    380 &   1.532 &\scriptsize    0.59  \\\hline
west0067   &  195 &\scriptsize    185 &   1.578 &\scriptsize    0.51  \\\hline
will57     &   55 &\scriptsize     17 &   0.531 &\scriptsize    0.05  \\\hline
can96      & 7787 &\scriptsize 259231 & 154.921 &\scriptsize 1131.60  \\\hline
steam3     &  108 &\scriptsize    313 &   1.828 &\scriptsize    0.81  \\\hline
curtis54   &   77 &\scriptsize    167 &   0.625 &\scriptsize    0.20  \\\hline
can73      & 1138 &\scriptsize   6997 &   8.266 &\scriptsize   21.82  \\\hline
bfw62a     &    7 &\scriptsize     11 &   0.579 &\scriptsize    0.05  \\\hline
ibm32      &   69 &\scriptsize     72 &   0.313 &\scriptsize    0.07  \\\hline
pores\_1   &   48 &\scriptsize      7 &   0.204 &\scriptsize    0.09  \\\hline
can61      &   90 &\scriptsize      5 &   2.109 &\scriptsize    0.82  \\\hline
bcsstk01   &    0 &\scriptsize     39 &   0.063 &\scriptsize    1.63  \\\hline
can24      &    0 &\scriptsize     17 &   0.000 &\scriptsize    0.05  \\\hline
fidapm05   &   44 &\scriptsize      5 &   0.422 &\scriptsize    0.10  \\\hline
fidap005   &   28 &\scriptsize      5 &   0.109 &\scriptsize    0.04  \\\hline \hline
\textbf{moyenne}&\textbf{467}&\textbf{\scriptsize 11219}&\textbf{7.867}&\textbf{\scriptsize 48.46} \\\hline
\end{tabular}
\end{table}

En comparant nos r\'esultats \`a ceux de Balas et De Souza sur les instances MM-I, nous obtenons :
\begin{itemize}
\item un nombre moyen de 467 noeuds contre 11219.
\item un temps moyen de r\'esolution de 7,867 secondes contre 48,46.
\end{itemize}
\newpage

Les graphes de la cat\'egorie MM-II ont au moins $104$ et au plus $147$ sommets. Ils ont \'et\'e obtenus \`a partir de matrices dont le nombre de colonnes varie entre $100$ et $200$. Le pr\'efixe \texttt{L125} est ajout\'e au nom de la matrice initiale lorsque le graphe est issu de la sous-matrice form\'ee de ses $125$ premi\`eres colonnes. La densit\'e de ces graphes varie de $0,05$ \`a $0,93$. 
\begin{table}[!h]
\small
\centering
\caption{R\'esultats pour les instances MatrixMarket MM\_II}
\label{resultmm2}
\begin{tabular}{|l||r|r|r|r|} 
\hline
Instance     & \textit{nb. noeuds} & \textit{\scriptsize nb. noeuds B.S.} & \textit{temps(s)} & \textit{\scriptsize temps(s) B.S.} \\ \hline \hline
L125.ash608     &   142 &\scriptsize     82 &   2.375 &\scriptsize    0.50 \\\hline
L125.will199    &   135 &\scriptsize    119 &   2.562 &\scriptsize    0.35 \\\hline
L125.west0167   &    69 &\scriptsize     73 &   1.906 &\scriptsize    0.22 \\\hline
ash331          &   183 &\scriptsize    212 &   2.812 &\scriptsize    0.82 \\\hline
west0132        &   105 &\scriptsize     85 &   4.297 &\scriptsize    0.42 \\\hline
rw136           &  1635 &\scriptsize  17386 &  28.875 &\scriptsize   57.57 \\\hline
bcspwr03        &    99 &\scriptsize     73 &   2.422 &\scriptsize    0.38 \\\hline
gre\_\_115      &  5573 &\scriptsize   6659 &  82.812 &\scriptsize   28.77 \\\hline
L125.dw\_\_162  &   469 &\scriptsize    214 &   8.360 &\scriptsize    1.15 \\\hline
L125.can\_\_187 &   786 &\scriptsize   5353 &  16.984 &\scriptsize   23.22 \\\hline
L125.gre\_\_185 &  1300 &\scriptsize   2337 &  30.578 &\scriptsize   31.40 \\\hline
L125.can\_\_161 & 15063 &\scriptsize 218627 & 580.797 &\scriptsize  (1800) $\star$ \\\hline
L125.lop163     &   308 &\scriptsize   6319 &  16.031 &\scriptsize   33.22 \\\hline
can\_\_144      &  1362 &\scriptsize  64203 &  31.250 &\scriptsize  443.49 \\\hline
lund\_a         &   385 &\scriptsize   2705 &  45.891 &\scriptsize  155.29 \\\hline
L125.bcsstk05   &   331 &\scriptsize   2301 &  30.812 &\scriptsize   54.29 \\\hline
L125.dwt\_\_193 &   259 &\scriptsize    129 &  29.719 &\scriptsize   73.74 \\\hline
L125.fs\_183\_1 &   240 &\scriptsize   1175 &  35.016 &\scriptsize   36.21 \\\hline
bcsstk04        &   600 &\scriptsize     89 &  26.796 &\scriptsize   80.01 \\\hline
arc130          &   268 &\scriptsize    129 &  99.328 &\scriptsize  137.42 \\\hline\hline
\textbf{moyenne}&\textbf{1466}&\textbf{\scriptsize 16414}&\textbf{53.981}&\textbf{\scriptsize 147.92}\\\hline
\end{tabular}
\end{table}
En comparant nos r\'esultats \`a ceux de Balas et De Souza sur les instances MM-II, nous obtenons :
\begin{itemize}
\item un nombre moyen de 1466 noeuds contre 16414.
\item un temps moyen de r\'esolution de 53,981 secondes contre 147,92.
\item toutes nos solutions sont optimales (en moins de 30 minutes) y compris pour \texttt{L125.can\_\_161} non r\'esolue par Balas et De Souza.
\end{itemize}
~\\~\\~\\~\\
Les graphes de la cat\'egorie MM-HD ont tous une densit\'e sup\'erieure \`a $0,35$. Ils contiennent $80$, $100$ ou $120$ sommets. Les pr\'efixes \texttt{L80, L100 et L120} sont ajout\'es au nom de la matrice initiale lorsque le graphe est issu de la sous-matrice form\'ee de ses $80$, $100$ ou $120$ premi\`eres colonnes. 
\newpage
\begin{table}[!h]
\small 	
\centering
\caption{R\'esultats pour les instances MatrixMarket MM-HD}
\label{resultdim}
\begin{tabular}{|l||r|r|r|r|r|r|} 
\hline
Instance     & \textit{nb. noeuds} & \textit{\scriptsize nb. noeuds B.S.} & \textit{temps(s)} & \textit{\scriptsize temps(s) B.S.} \\ \hline \hline
L100.steam2	    &	175	&	\scriptsize 167	&	8.781	&	\scriptsize 30.98	\\\hline
L100.cavity01	&	115	&	\scriptsize 7	&	10.657	&	\scriptsize 3.52	\\\hline
L80.cavity01	&	125	&	\scriptsize 15	&	4.578	&	\scriptsize 2.22	\\\hline
L80.fidap025	&	52	&	\scriptsize 7	&	2.625	&	\scriptsize 1.14	\\\hline
L120.fidap025	&	82	&	\scriptsize 21	&	9.61	&	\scriptsize 9.27	\\\hline
L80.steam2	    &	115	&	\scriptsize 29	&	4.218	&	\scriptsize 3.15	\\\hline
L100.fidap021	&	119	&	\scriptsize 5	&	5.469	&	\scriptsize 2.64	\\\hline
L100.fidap025	&	310	&	\scriptsize 11	&	13.578	&	\scriptsize 5.41	\\\hline
L120.cavity01	&	138	&	\scriptsize 15	&	16.375	&	\scriptsize 16.57	\\\hline
L120.fidap021	&	192	&	\scriptsize 27	&	16.765	&	\scriptsize 17.36	\\\hline
L80.fidap021	&	99	&	\scriptsize 7	&	4.031	&	\scriptsize 2.43	\\\hline
L120.rbs480a	&	209	&	\scriptsize 233	&	21.282	&	\scriptsize 98.84	\\\hline
L120.wm2	    &	171	&	\scriptsize 33	&	18.781	&	\scriptsize 19.27	\\\hline
L100.rbs480a	&	156	&	\scriptsize 63	&	13.953	&	\scriptsize 11.91	\\\hline
L80.wm3	        &	184	&	\scriptsize 13	&	8.406	&	\scriptsize 3.81	\\\hline
L80.wm1	        &	155	&	\scriptsize 59	&	8.64	&	\scriptsize 11.13	\\\hline
L80.rbs480a	    &	43	&	\scriptsize 5	&	3.5	    &	\scriptsize 0.97	\\\hline
L80.wm2	        &	153	&	\scriptsize 15	&	7.36	&	\scriptsize 2.89	\\\hline
L100.wm3	    &	179	&	\scriptsize 9	&	19.656	&	\scriptsize 4.79	\\\hline
L120.e05r0000	&	82	&	\scriptsize 11	&	9.828	&	\scriptsize 7.51	\\\hline
L100.wm1	    &	196	&	\scriptsize 27	&	27.782	&	\scriptsize 11.65	\\\hline
L120.fidap022	&	219	&	\scriptsize 177	&	30.782	&	\scriptsize 94.06	\\\hline
L100.wm2	    &	190	&	\scriptsize 15	&	18.485	&	\scriptsize 7.37	\\\hline
L100.fidapm02	&	255	&	\scriptsize 13	&	25.125	&	\scriptsize 6.35	\\\hline
L120.fidap001	&	237	&	\scriptsize 27	&	37.532	&	\scriptsize 22.56	\\\hline
L100.e05r0000	&	120	&	\scriptsize 47	&	7.937	&	\scriptsize 14.94	\\\hline
L120.fidapm02	&	223	&	\scriptsize 11	&	42.953	&	\scriptsize 12.29	\\\hline
L80.fidapm02	&	50	&	\scriptsize 7	&	3.281	&	\scriptsize 1.97	\\\hline
L100.fidap001	&	100	&	\scriptsize 39	&	5.141	&	\scriptsize 9.82	\\\hline
L100.fidap022	&	168	&	\scriptsize 143	&	13.985	&	\scriptsize 30.62	\\\hline
L80.e05r0000	&	133	&	\scriptsize 3	&	7.937	&	\scriptsize 0.96	\\\hline
L80.fidap001	&	84	&	\scriptsize 1	&	3.14	&	\scriptsize 0.98	\\\hline
L80.fidap022	&	75	&	\scriptsize 171	&	2.172	&	\scriptsize 10.05	\\\hline
L80.fidap002	&	83	&	\scriptsize 9	&	3.328	&	\scriptsize 1.71	\\\hline
L80.fidap027	&	81	&	\scriptsize 3	&	4.469	&	\scriptsize 1.18	\\\hline
L100.fidap027	&	187	&	\scriptsize 5	&	23.437	&	\scriptsize 4.15	\\\hline
L100.fidap002	&	102	&	\scriptsize 3	&	6.234	&	\scriptsize 2.38	\\\hline
L120.fidap002	&	121	&	\scriptsize 71	&	10.453	&	\scriptsize 37.86	\\\hline
L120.fidap027	&	229	&	\scriptsize 9	&	50.719	&	\scriptsize 9.07	\\\hline\hline
\textbf{moyenne}&\textbf{146}&\textbf{\scriptsize 39}&\textbf{13.666}&\textbf{\scriptsize 13.74}\\\hline
\end{tabular}
\end{table}
En comparant nos r\'esultats \`a ceux de Balas et De Souza sur les instances \hbox{MM-HD}, nous obtenons un nombre moyen de 146 noeuds contre 39 et un temps moyen de r\'esolution de 13,666 secondes contre 13,74.
\newpage
\newpage

\subsection{Synth\`ese}
Il semble \`a l'issue de ces tests que notre algorithme soit plus performant que celui de Balas et De Souza sur les instances difficiles. Ses performances sont directement li\'ees \`a la structure du graphe initial. En effet, la s\'eparation d'un graphe dans lequel toute paire de sommets non adjacents est associ\'ee \`a un nombre \'elev\'e de cha\^ines sommets-disjoints est extr\`emement rapide pour les instances test\'ees. Les exemples de DSJC125.5, DSJC125.9, queen10\_10, queen9\_9 mettent en \'evidence ce constat. Sur ces quatre instances pour lesquelles $\alpha_{min}$ est \'elev\'e, nous explorons en moyenne 1623 noeuds contre 117314 pour Balas et De Souza. Notre temps de r\'esolution moyen est de 32,363 secondes et toutes nos solutions sont optimales. Balas et De Souza r\'esolvent seulement deux de ces instances \`a l'optimal en 930,93 secondes en moyenne.
La densit\'e du graphe initial affecte les performances de l'algorithme de Balas et De Souza. Notre algorithme est pour sa part mieux adapt\'e au traitement des graphes denses, susceptibles d'\^etre dot\'es de valeurs \'elev\'ees pour $\alpha_{min}$.  
\section{Conclusions et perspectives}
Dans ce rapport, nous \'etudions une formulation du probl\`eme du s\'eparateur d'un graphe $G=(V,E)$ connexe non orient\'e gr\^ace \`a un programme lin\'eaire mixte, donn\'e par Balas et De Souza, que nous avons enrichi. Nos travaux nous ont permis d'obtenir la dimension du poly\`edre $P_{ab}$ associ\'e aux ab-s\'eparateurs du graphe, enveloppe convexe de l'ensemble des vecteurs d'incidence des partitions de V v\'erifiant le VSP avec a et b fix\'es. Nous avons introduit de nouvelles in\'egalit\'es valides en consid\'erant les cha\^ines sommets-disjoints entre a et b. L'algorithme que nous avons impl\'ement\'e utilise ces in\'egalit\'es et nous permet d'am\'eliorer sensiblement les r\'esultats obtenus par Balas et De Souza \cite{Bal05b}.\\
Nous proposons parall\`element un mod\`ele introduisant des variables sur les ar\^etes du graphe.

Nous poursuivons actuellement notre \'etude du VSP dans une direction th\'eorique et dans une direction exp\'erimentale. Du point de vue th\'eorique, l'approche poly\'edrale du probl\`eme nous conduit \`a rechercher les facettes d\'efinies par les in\'egalit\'es valides de notre mod\'elisation. Nous souhaitons \'egalement compl\'eter le mod\`ele utilisant des variables sur les sommets par l'ajout de celui utilisant des variables sur les ar\^etes pour poursuivre la recherche de nouvelles in\'egalit\'es valides am\'eliorant la description poly\'edrale du probl\`eme. Du point de vue exp\'erimental, nous travaillons \`a la mise au point de strat\'egies de s\'eparation des in\'egalit\'es que nous avons introduites. Nous pr\'eparons \'egalement une s\'erie de tests qui nous permettra d'\'evaluer l'efficacit\'e de notre algorithme sur des instances de grandes tailles.
%
\newpage
\bibliographystyle{plain}
\bibliography{vspmastermj}
\newpage
\pagebreak
\appendix

\noindent \textbf{\large Annexes :}\\

\noindent Ces annexes rassemblent les caract\'eristiques des graphes qui composent les s\'eries d'instances sur lesquelles nous avons travaill\'e. Dans chaque tableau, $n$ est le nombre de sommets, $e$ est le nombre d'ar\^etes, $d$ est la densit\'e du graphe ($d=\frac{2e}{n(n-1)}$), $\beta(n)$ est la borne sur le cardinal des ensembles $A$ et $B$ ($\beta(n)=\frac{2n}{3}$), $\alpha_{min}$ est le nombre minimum de sommets appartenant au s\'eparateur $C$.

\section{Instances DIMACS}
\begin{table}[!h]
\small
\centering
\caption{Instances DIMACS}
\label{instdim}
\begin{tabular}{|l||r|r|r|r|r|r|} 
\hline
Instance     & $n$ &  $e$  & $d$  & $\beta(n)$ & $\alpha_{min}$& $\alpha_{min} / n$  \\ \hline \hline
david        & 87  &  406  & 0.11 &     58     &     1         & 0.011 \\\hline
DSJC125.1    & 125 &  736  & 0.09 &     83     &     5         & 0.040 \\\hline
DSJC125.5    & 125 &  3891 & 0.50 &     83     &     51        & 0.408 \\\hline
DSJC125.9    & 125 &  6961 & 0.90 &     83     &     103       & 0.824 \\\hline
games120     & 120 &  638  & 0.09 &     80     &     2         & 0.017 \\\hline
miles500     & 128 &  1170 & 0.14 &     85     &     2         & 0.016 \\\hline
miles750     & 128 &  2213 & 0.26 &     85     &     6         & 0.047 \\\hline
miles1000    & 128 &  3216 & 0.40 &     85     &     11        & 0.086 \\\hline
myciel3      & 11  &  20   & 0.36 &     7      &     3         & 0.273 \\\hline
myciel4      & 23  &  71   & 0.28 &     15     &     4         & 0.174 \\\hline
myciel5      & 47  &  236  & 0.22 &     31     &     5         & 0.106 \\\hline
myciel6      & 95  &  755  & 0.17 &     63     &     6         & 0.063 \\\hline
myciel7      & 191 &  2360 & 0.13 &     127    &     7         & 0.037 \\\hline
queen6\_ 6   & 36  &  290  & 0.46 &     24     &     15        & 0.417 \\\hline
queen7\_ 7   & 49  &  476  & 0.40 &     32     &     18        & 0.367 \\\hline
queen8\_ 8   & 64  &  728  & 0.36 &     42     &     21        & 0.328 \\\hline
queen8\_ 12  & 96  &  1368 & 0.30 &     64     &     25        & 0.260 \\\hline
queen9\_ 9   & 81  &  1056 & 0.33 &     54     &     24        & 0.296 \\\hline
queen10\_ 10 & 100 &  1470 & 0.30 &     66     &     27        & 0.270 \\\hline
queen11\_ 11 & 121 &  1980 & 0.27 &     80     &     30        & 0.248 \\\hline
queen12\_ 12 & 144 &  2596 & 0.25 &     96     &     33        & 0.229 \\\hline
\end{tabular}
\end{table}

\pagebreak

\section{Instances MatrixMarket}
\begin{table}[!h]
\small
	\centering
\caption{Instances MatrixMarket MM-I}
\label{instmm1}
\begin{tabular}{|l||r|r|r|r|r|r|} 
\hline
Instance     & $n$ &  $e$  & $d$  & $\beta(n)$ & $\alpha_{min}$& $\alpha_{min} / n$  \\ \hline \hline
ash219       & 85  &  219  & 0.06 &     56     &     2         & 0.024 \\\hline
dwt72        & 72  &  170  & 0.07 &     48     &     2         & 0.028 \\\hline
can62        & 62  &  210  & 0.11 &     41     &     2         & 0.032 \\\hline
dwt66        & 66  &  255  & 0.12 &     44     &     4         & 0.061 \\\hline
bcspwr02     & 49  &  177  & 0.15 &     32     &     2         & 0.041 \\\hline
dwt\_\_59    & 59  &  256  & 0.15 &     39     &     3         & 0.051 \\\hline
bcspwr01     & 39  &  118  & 0.16 &     26     &     2         & 0.051 \\\hline
ash85        & 85  &  616  & 0.17 &     56     &     6         & 0.071 \\\hline
dwt87        & 87  &  726  & 0.19 &     58     &     4         & 0.046 \\\hline
impcol\_b    & 59  &  329  & 0.19 &     39     &     3         & 0.051 \\\hline
west0067     & 67  &  411  & 0.19 &     44     &     3         & 0.045 \\\hline
will57       & 57  &  304  & 0.19 &     38     &     2         & 0.035 \\\hline
can96        & 96  &  912  & 0.20 &     64     &     14        & 0.146 \\\hline
steam3       & 80  &  712  & 0.23 &     53     &     8         & 0.100 \\\hline
curtis54     & 54  &  337  & 0.24 &     36     &     5         & 0.093 \\\hline
can73        & 73  &  652  & 0.25 &     48     &     10        & 0.137 \\\hline
bfw62a       & 62  &  639  & 0.34 &     41     &     3         & 0.048 \\\hline
ibm32        & 32  &  179  & 0.36 &     21     &     4         & 0.125 \\\hline
pores\_1     & 30  &  179  & 0.41 &     20     &     5         & 0.167 \\\hline
can61        & 61  &  866  & 0.47 &     40     &     12        & 0.197 \\\hline
bcsstk01     & 48  &  622  & 0.55 &     32     &     18        & 0.375 \\\hline
can24        & 24  &  156  & 0.57 &     16     &     8         & 0.333 \\\hline
fidapm05     & 42  &  528  & 0.61 &     28     &     12        & 0.286 \\\hline
fidap005     & 27  &  234  & 0.67 &     18     &     9         & 0.333 \\\hline
\end{tabular}
\end{table}

\begin{table}[!h]
\small
	\centering
\caption{Instances MatrixMarket MM\_II}
\label{instmm2}
\begin{tabular}{|l||r|r|r|r|r|r|} 
\hline
Instance        & $n$ &  $e$  & $d$  & $\beta(n)$ & $\alpha_{min}$ & $\alpha_{min} / n$ \\ \hline \hline
L125.ash608     & 125 &  390 & 0.05 &   83        &     2          & 0.016 \\\hline
L125.will199    & 125 &  386 & 0.05 &   83        &     1          & 0.008 \\\hline
L125.west0167   & 125 &  444 & 0.06 &   83        &     1          & 0.008 \\\hline
ash331          & 104 &  331 & 0.06 &   69        &     2          & 0.019 \\\hline
west0132        & 132 &  560 & 0.06 &   88        &     1          & 0.008 \\\hline
rw136           & 136 &  641 & 0.07 &   90        &     1          & 0.007 \\\hline
bcspwr03        & 118 &  576 & 0.08 &   78        &     2          & 0.017 \\\hline
gre\_\_115      & 115 &  576 & 0.09 &   76        &     3          & 0.026 \\\hline
L125.dw\_\_162  & 125 &  943 & 0.12 &   83        &     6          & 0.048 \\\hline
L125.can\_\_187 & 125 & 1022 & 0.13 &   83        &     12         & 0.096 \\\hline
L125.gre\_\_185 & 125 & 1177 & 0.15 &   83        &     6          & 0.048 \\\hline
L125.can\_\_161 & 125 & 1257 & 0.16 &   83        &     12         & 0.096 \\\hline
L125.lop163     & 125 & 1218 & 0.16 &   83        &     9          & 0.072 \\\hline
can\_\_144      & 144 & 1656 & 0.16 &   96        &     18         & 0.125 \\\hline
lund\_a         & 147 & 2837 & 0.26 &   98        &     11         & 0.075 \\\hline
L125.bcsstk05   & 125 & 2701 & 0.35 &   83        &     17         & 0.136 \\\hline
L125.dwt\_\_193 & 125 & 2982 & 0.38 &   83        &     21         & 0.168 \\\hline
L125.fs\_183\_1 & 125 & 3392 & 0.44 &   83        &     1          & 0.008 \\\hline
bcsstk04        & 132 & 5918 & 0.68 &   88        &     48         & 0.364 \\\hline
arc130          & 130 & 7763 & 0.93 &   86        &     21         & 0.162 \\\hline
\end{tabular}
\end{table}

\begin{table}[!h]
\small
	\centering
\caption{Instances MatrixMarket MM-HD}
\label{instmmhd}
\begin{tabular}{|l||r|r|r|r|r|r|} 
\hline
Instance       & $n$ &  $e$ &  $d$ &  $\beta(n)$ & $\alpha_{min}$ & $\alpha_{min} / n$  \\ \hline \hline
L100.steam2	   & 100 & 1766	& 0.36 &	66	     &	16	          & 0.160 \\\hline
L100.cavity01  & 100 & 1844	& 0.37 &	66	     &	8	          & 0.080 \\\hline
L80.cavity01   &  80 & 1201	& 0.38 &	53	     &	8	          & 0.100 \\\hline
L80.fidap025   &  80 & 1201	& 0.38 &	53	     & 	12	          & 0.150 \\\hline
L120.fidap025  & 120 & 2787	& 0.39 &	80	     &	18	          & 0.150 \\\hline
L80.steam2     &  80 & 1272	& 0.40 &	53	     &	12	          & 0.150 \\\hline
L100.fidap021  & 100 & 2028	& 0.41 &	66	     &	15	          & 0.150 \\\hline
L100.fidap025  & 100 & 2031	& 0.41 &	66	     &	12	          & 0.120 \\\hline
L120.cavity01  & 120 & 2972	& 0.42 &	80	     &	8	          & 0.067 \\\hline
L120.fidap021  & 120 & 3058	& 0.43 &	80	     &	15	          & 0.125 \\\hline
L80.fidap021   &  80 & 1347	& 0.43 &	53	     &	11	          & 0.138 \\\hline
L120.rbs480a   & 120 & 3273	& 0.46 &	80	     &	23	          & 0.192 \\\hline
L120.wm2       & 120 & 3387	& 0.47 &	80	     &	1	          & 0.008 \\\hline
L100.rbs480a   & 100 & 2550	& 0.52 &	66	     &	26	          & 0.260 \\\hline
L80.wm3        &  80 & 1739	& 0.55 &	53	     &	1	          & 0.013 \\\hline
L80.wm1        &  80 & 1786	& 0.57 &	53	     &	2	          & 0.025 \\\hline
L80.rbs480a    &  80 & 1819	& 0.58 &	53       &	18	          & 0.225 \\\hline
L80.wm2        &  80 & 1848	& 0.58 &	53	     &	1	          & 0.013 \\\hline
L100.wm3       & 100 & 2934	& 0.59 &	66	     &	1	          & 0.010 \\\hline
L120.e05r0000  & 120 & 4177	& 0.59 &	80	     &	30	          & 0.250 \\\hline
L100.wm1       & 100 & 2956	& 0.60 &	66	     &	1	          & 0.010 \\\hline
L120.fidap022  & 120 & 4307	& 0.60 &	80	     &	30	          & 0.250 \\\hline
L100.wm2       & 100 & 3039	& 0.61 &	66	     &	1	          & 0.010 \\\hline
L100.fidapm02  & 100 & 3090	& 0.62 &	66	     &	27	          & 0.270 \\\hline
L120.fidap001  & 120 & 4482	& 0.63 &	80	     &	35         	  & 0.292 \\\hline
L100.e05r0000  & 100 & 3153	& 0.64 &	66	     &	30	          & 0.300 \\\hline
L120.fidapm02  & 120 & 4626	& 0.65 &	80	     &	27	          & 0.225 \\\hline
L80.fidapm02   &  80 & 2058	& 0.65 &	53	     &	27	          & 0.338 \\\hline
L100.fidap001  & 100 & 3379	& 0.68 &	66	     &	36	          & 0.360 \\\hline
L100.fidap022  & 100 & 3359	& 0.68 &	66	     &	38	          & 0.380 \\\hline
L80.e05r0000   &  80 & 2150	& 0.68 &	53	     &	19	          & 0.238 \\\hline
L80.fidap001   &  80 & 2288	& 0.72 &	53	     &	26	          & 0.325 \\\hline
L80.fidap022   &  80 & 2403	& 0.76 &	53	     &	39	          & 0.488 \\\hline
L80.fidap002   &  80 & 2422	& 0.77 &	53	     &	27	          & 0.338 \\\hline
L80.fidap027   &  80 & 2518	& 0.80 &	53	     &	24	          & 0.300 \\\hline
L100.fidap027  & 100 & 4014	& 0.81 &	66	     &	27	          & 0.270 \\\hline
L100.fidap002  & 100 & 4057	& 0.82 &	66	     &	34	          & 0.340 \\\hline
L120.fidap002  & 120 & 5886	& 0.82 &	80	     &	52	          & 0.433 \\\hline
L120.fidap027  & 120 & 6051	& 0.85 &	80	     &	35	          & 0.292 \\\hline
\end{tabular}
\end{table}

\end{document}